\documentclass[proof]{WileyASNA-v1}
\articletype{Original article}%

\usepackage{multirow}

\received{DD Month 2019}
\revised{DD Month YYYY}
\accepted{DD Month YYYY}

\begin{document}

\title{Estimation of the detected background by the future gamma-ray transient mission CAMELOT}
\author[1,2,3]{J. {\v R}{\'i}pa*}
\author[2]{G. Galg\'oczi}
\author[1,4,5]{N. Werner}
\author[6]{A. P\'al}
\author[1,2,5,6]{M. Ohno}
\author[6]{L. M\'esz\'aros}
\author[7]{T. Mizuno}
\author[8]{N. Tarcai}
\author[5]{K. Torigoe}
\author[5]{N. Uchida}
\author[5]{Y. Fukazawa}
\author[5]{H. Takahashi}
\author[9]{K. Nakazawa}
\author[5]{N. Hirade}
\author[5]{K. Hirose}
\author[9]{S. Hisadomi}
\author[10]{T. Enoto}
\author[11]{H. Odaka}
\author[12]{Y. Ichinohe}
\author[2]{Z. Frei}
\author[6]{L. Kiss}

\authormark{{\v R}{\'I}PA \textsc{et al}}

\address[1]{\orgname{MTA-E\"ot\"vos University Lend\"ulet Hot Universe Research Group}, \orgaddress{\state{Budapest}, \country{Hungary}}}

\address[2]{\orgdiv{Institute of Physics}, \orgname{E\"otv\"os University}, \orgaddress{\state{Budapest}, \country{Hungary}}}

\address[3]{\orgdiv{Astronomical Institute}, \orgname{Charles University}, \orgaddress{\state{Prague}, \country{Czech Republic}}}

\address[4]{\orgdiv{Department of Theoretical Physics and Astrophysics, Faculty of Science}, \orgname{Masaryk University}, \orgaddress{\state{Brno}, \country{Czech Republic}}}

\address[5]{\orgdiv{School of Science}, \orgname{Hiroshima University}, \orgaddress{\state{Higashi-Hiroshima}, \country{Japan}}}

\address[6]{\orgname{Konkoly Observatory of the Hungarian Academy of Sciences}, \orgaddress{\state{Budapest}, \country{Hungary}}}

\address[7]{\orgdiv{Hiroshima Astrophysical Science Center}, \orgname{Hiroshima University}, \orgaddress{\state{Higashi-Hiroshima}, \country{Japan}}}

\address[8]{\orgname{C3S Electronics Development LLC.}, \orgaddress{\state{Budapest}, \country{Hungary}}}

\address[9]{\orgdiv{Department of Physics}, \orgname{Nagoya University}, \orgaddress{\state{Nagoya}, \country{Japan}}}

\address[10]{\orgdiv{The Hakubi Center for Advanced Research and Department of Astronomy}, \orgname{Kyoto University}, \orgaddress{\state{Kyoto}, \country{Japan}}}

\address[11]{\orgdiv{$^{11}$Department of Physics}, \orgname{University of Tokyo}, \orgaddress{\state{Tokyo}, \country{Japan}}}

\address[12]{\orgdiv{Department of Physics, Rikkyo University}, \orgname{Rikkyo University}, \orgaddress{\state{Tokyo}, \country{Japan}}}

\corres{*J. {\v R}{\'i}pa, MTA-E\"otv\"os University Lend\"ulet Hot Universe Research Group, P\'azm\'any P\'eter s\'et\'any 1/A, Budapest, 1117, Hungary. \email{jripa@caesar.elte.hu}}

\abstract{This study presents a background estimation for the Cubesats Applied for MEasuring and LOcalising Transients (CAMELOT), which is a proposed fleet of nanosatellites for the all-sky monitoring and timing based localization of gamma-ray transients with precise localization capability at low Earth orbits. CAMELOT will allow to observe and precisely localize short gamma-ray bursts (GRBs) associated with kilonovae, long GRBs associated with core-collapse massive stars, magnetar outbursts, terrestrial gamma-ray flashes, and gamma-ray counterparts to gravitational wave sources. The fleet of at least nine 3U CubeSats is proposed to be equipped with large and thin CsI(Tl) scintillators read out by multi-pixel photon counters (MPPC). A careful study of the radiation environment in space is necessary to optimize the detector ca\-sing, estimate the duty cycle due to the crossing of the South Atlantic Anomaly and polar regions, and to minimize the effect of the radiation damage of MPPCs.}


\jnlcitation{\cname{%
\author{{\v R}{\'i}pa J.}, 
\author{Galg\'oczi G.}, 
\author{Werner N.}, et al.}, 
\ctitle{Estimation of the detected background by the future gamma-ray transient mission CAMELOT}, \cjournal{Astron. Nachr.}, \cvol{YYYY;VVV:PPP--PPP}.}

\maketitle

\section{Introduction}\label{sec:intro}
Cubesats Applied for MEasuring and LOcalising Transients (\textit{CAMELOT}), for details see \cite{Ohno2018, Pal2018, Ripa2018, Werner2018, Torigoe2019}, is a future constellation of at least nine 3U CubeSats, which will be primarily monitoring and precisely and rapidly localizing gamma-ray bursts (GRBs), see \cite{Klebesadel1973, Vedrenne2009, Kouveliotou2012}, over the whole sky. It will allow a regular detection of electromagnetic counterparts of gravitation wave sources si\-mi\-lar to the breakthrough discovery of the neutron star merger GW170817 \citep{Abbott2017} associated with a short GRB 170817A \citep{Goldstein2017}. This gamma-ray monitoring network will also allow to observe long GRBs associated with core-collapse massive stars, soft gamma repeters (SGR) associated with magnetar outbursts \citep{Mazets1979, Kouveliotou1998}, and terrestrial gamma-ray flashes (TGF) produced by thunderstorms \citep{Fishman1994}.

\begin{figure}[t]
	\centerline{\includegraphics[width=0.95\linewidth]{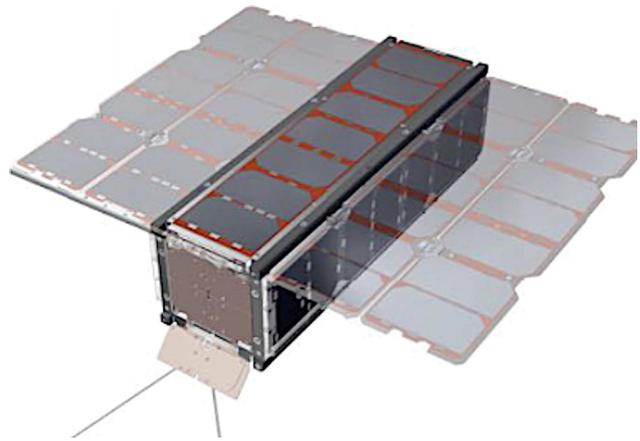}}
	\caption{A schematic of one satellite of the \textit{CAMELOT} constellation which can use a 3U CubeSat platform developed by C3S LLC for the RadCube mission.\label{fig:camelot}}
\end{figure}

The \textit{CAMELOT} satellites (see Figure~\ref{fig:camelot}) will be equipped with two or four large and thin CsI(Tl) scintillators, $75\times150\times5$~mm$^3$ each, read out by Multi-Pixel Photon Counter (MPPC) silicon photomultipliers from Hamamatsu. The scintillators will be enclosed in Al or carbon fiber reinforced plastic (CFRP) support structure and will be placed on one or two of the long sides of the satellite with sensitivity in the range of $\sim10-1000$ keV.

The intended orbit configurations of the \textit{CAMELOT} CubeSats are Low Earth Orbits (LEO) at altitude $\sim500-600$ km either with inclination of $53^\circ$ or polar orbits with inclination around $97.6^\circ$. However, final decision will depend on the lunch opportunities.

Here we briefly summarize our studies of the impact of the radiation environment in LEO on the \textit{CAMELOT} satellites carried out to optimize the scintillator casing, estimate the duty cycle due to crossing of the South Atlantic Anomaly (SAA) and polar regions and to design radiation shielding to minimize the effect of the degradation of MPPCs due to the proton fluxes.

\section{Background components}\label{sec:background}
The aim of this work is to estimate the total background count rate due to the space environment in the part of the orbit ideal for gamma-ray transient observations, i.e. outside SAA and polar regions which contain high fluxes of geomagnetically trapped particles. Spectra of various background components are used as an input in the full Monte Carlo (MC) simulation applied in Geant4\footnote{\url{https://geant4.web.cern.ch}} \citep{Agostinelli2003,Allison2016} together with the satellite's mass model.

\subsection{Trapped particles}\label{sec:trapped}
For the fluxes of the trapped e$^-$ and p$^+$ the AE9 and AP9 mo\-dels \citep{Ginet2013} implemented in ESA's SPace ENVironment Information System (SPENVIS\footnote{\url{www.spenvis.oma.be}}) were used, respectively. AE9/AP9 are based on 33 satellite datasets from 1976 to 2011 developed by U.S. Air Force Research \footnote{\url{https://www.vdl.afrl.af.mil/programs/ae9ap9}}. The parameters of these models were set to MC mode with 100 runs and mean aggregate.

\begin{figure}[t]
\	\centerline{\includegraphics[width=\linewidth]{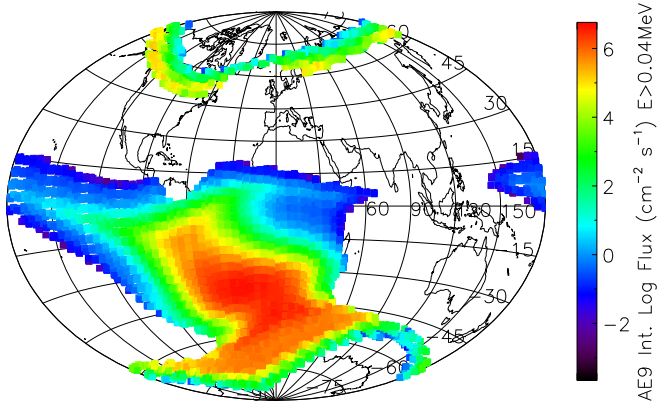}}
	\caption{A map of the integral flux of geomagnetically trapped electrons at 500~km altitude according to the AE9 model (MC model, 50~\% CL).\label{fig:AE9}}
\end{figure}

\begin{figure}[t]
	\centerline{\includegraphics[width=\linewidth]{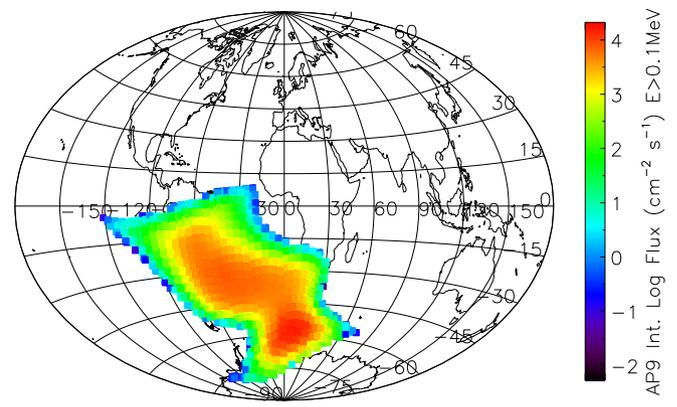}}
	\caption{A map of the integral flux of geomagnetically trapped protons at 500~km altitude according to the AP9 model (MC model, 50~\% CL).\label{fig:AP9}}
\end{figure}

Averaged fluxes of e$^-$ and p$^+$ over three circular orbits (avoiding crossing SAA) at an altitude of 500~km outside SAA and polar regions with inclination $i=20^\circ$ were simulated in SPENVIS. Figure~\ref{fig:AE9} and Figure~\ref{fig:AP9} show maps of trapped e$^-$ and p$^+$, respectively. The flux of trapped p$^+$ outside SAA is negligible. The orbit averaged differential spectrum of trapped e$^-$ is shown in Figure~\ref{fig:bkg_spectra}. The integral flux ($E>40$~keV) is 2.0~cm$^{-2}$s$^{-1}$.

\subsection{Cosmic X-ray background}
The isotropic cosmic X-ray background (CXB), see \cite{Giacconi1962, Boldt1981, Fabian1985, Mizuno2004, Campana2013}, can originate from the summation of the emi\-ssion of extragalactic sources (active galactic nuclei, quasi-stellar objects, supernovae Ia, galaxy clusters, starburst galaxies, hot intergalactic gas), details in \cite{Meszaros1987, Meszaros1988, Bagoly1988, Bi1990, Bi1991, Jahoda1991, Shanks1991, Sreekumar1998, Dean2003, Ajello2008, Ma2018}. Another origin discussed in li\-te\-ra\-ture is due to Cosmic Microwave Background \citep{Penzias1965} inverse Compton scattered on cosmic ray electrons, see \citet{Dean2003} and references therein. For our simulations we used the model by \citet{Gruber1999} which mo\-dels low energy and high energy part of the CXB measurements from \textit{HEAO-1} \citep{Rothschild1979}, \textit{CGRO}/COMPTEL and \textit{CGRO/EGRET} \citep{Gehrels1993} instruments over the wide range of energies from 3~keV to 100~GeV. This model is used as a standard in modeling of CXB for space missions.

The flux is omnidirectional and for 500~km altitude it irradiates the satellite from the solid angle of 8.64 srad (3.93 srad is occulted by the Earth). The CXB spectrum used for the Geant4 simulations is shown in Figure~\ref{fig:bkg_spectra}. The integral flux ($E>10$~keV) is 30.3~cm$^{-2}$s$^{-1}$.

\subsection{Galactic cosmic rays}
The spectra of primary particles of the galactic cosmic rays (GCR) used in the Geant4 simulations are described below and shown in Figure~\ref{fig:bkg_spectra}. For 500~km altitude the fluxes irradiate the satellite from the solid angle of 8.64 srad.

\subsubsection{Primary H and He}\label{sec:GCR_H_He}
For the spectra of the primary H and He we used the model ISO-15390\footnote{\url{www.iso.org/standard/37095.html}}, implemented in SPENVIS. ISO-15390 is the International Standard for estimating the radiation impact of galactic cosmic rays on hardware in space.

The averaged spectra were obtained for the same three orbits as described in Section~\ref{sec:trapped} with following pa\-ra\-me\-ters: solar minimum activity (May 1996), magnetic shielding on, stormy magnetosphere, St\o{}rmer with eccentric dipole method and magnetic field moment unchanged.

The integral flux for H is 0.12~cm$^{-2}$s$^{-1}$ ($E>1$~GeV) and for He it is 0.021~cm$^{-2}$s$^{-1}$ ($E>1$~GeV/n).

\subsubsection{Primary electrons and positrons}
For the spectra of the primary e$^-$ and e$^+$ we used the model described by \citet{Mizuno2004} (see also references therein) for solar minimum (solar modulation potential $\varphi=0.55$~GV), 500~km altitude and for geomagnetic latitude $\theta_\textrm{M}=20^\circ$.

The integral flux ($E>1$~GeV) for e$^-$ is $3\times10^{-3}$~cm$^{-2}$s$^{-1}$ and for e$^+$ it is $2.3\times10^{-4}$~cm$^{-2}$s$^{-1}$.

\subsection{Secondary particles and radiation}
\begin{figure}[t]
	\centerline{\includegraphics[width=\linewidth]{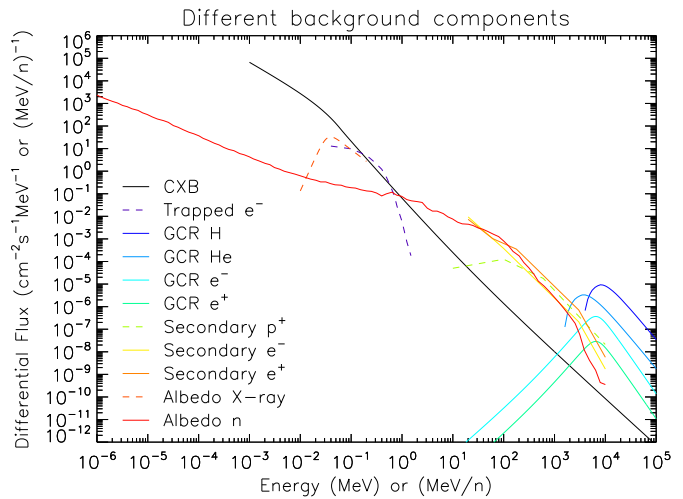}}
	\caption{Spectra of all background components (particles and X-rays) used to irradiate the mass model of the satellite in the Geant4 simulations. \label{fig:bkg_spectra}}
\end{figure}

Secondary (albedo) particles and radiation originates from the interaction of GCR with the Earth's atmosphere \citep{Jursa1985}. The fluxes used in the Geant4 simulations are described below and shown in Figure~\ref{fig:bkg_spectra}.

\subsubsection{Secondary protons}\label{sec:secondary_p}
For the secondary p$^+$ and for energy above 100~MeV we use the modeling by \citep{Mizuno2004} based on the Alpha Magnetic Spectrometer (AMS) data \citep{Alcaraz2000a} from 380~km altitude for the geomagnetic latitude $17^\circ<\theta_\textrm{M}<23^\circ$. For energy below 100~MeV we use the fit to \textit{MITA}/NINA-2 data from 450~km altitude \citep{Bidoli2002} for $1.0 \leq \textrm{L-shell} \leq 1.7$. For details see the LAT Technical Note LAT-TD-08316-01\footnote{\url{fermi.gsfc.nasa.gov/science/resources/swg/LAT_bkgd_Rev.pdf}} of the \textit{Fermi} satellite \citep{Atwood2009}.

There is only a small dependence of the flux on altitude \citep{Bidoli2002,Zuccon2003} therefore it can be used as an approximation to the flux at 500~km. The same flux model is used for upward and downward component therefore the flux irradiates the satellite from the solid angle of $4\pi$~srad. The integral flux ($E>10$~MeV) is 0.037~cm$^{-2}$s$^{-1}$.

\subsubsection{Secondary electrons and positrons}
For the secondary e$^-$ and e$^+$ and for energy above 100~MeV we use the modeling by \citep{Mizuno2004} based on the AMS data \citep{Alcaraz2000b} from 380~km altitude for the geomagnetic latitude $0^\circ<\theta_\textrm{M}<17^\circ$. For energy below 100~MeV we use the fit to the \textit{Mir}/MARIA-2 data from 400~km altitude \citep{Voronov1991} for $1.0 \leq \textrm{L-shell} \leq 1.2$. For details see the LAT Technical Note LAT-TD-08316-01.

The same flux model is used for upward and downward component therefore the flux irradiates the satellite from the solid angle of $4\pi$~srad. The integral flux ($E>20$~MeV) is 0.18~cm$^{-2}$s$^{-1}$ for e$^-$ and 0.23~cm$^{-2}$s$^{-1}$ for e$^+$.

\subsubsection{Albedo X-rays}
The spectrum of the albedo X-rays is taken from the \textit{Swift}/BAT measurements for altitude of $\sim 550$~km and inclination $i=20.6^\circ$ \citep{Ajello2008}. Concerning the altitude 500~km the flux would irradiate the satellite from the solid angle of 3.93~srad. The integral flux ($E>10$~keV) is 2.0~cm$^{-2}$s$^{-1}$.

\subsubsection{Albedo neutrons}
For the albedo neutrons we use the predictions of the QinetiQ Atmospheric Radiation Model (QARM), based on MC radiation transport code, as reported in the ESA document ECSS-E-ST-10-04C\footnote{\url{https://ecss.nl/standard/ecss-e-st-10-04c-space-environment}}. The fluxes are scaled from 100~km to 500~km as described in the document and are for the cutoff rigidity of 5 GV. The integral flux ($E>1$~eV) is 0.61~cm$^{-2}$s$^{-1}$.

\begin{figure}[t]
	\centerline{\includegraphics[width=0.8\linewidth]{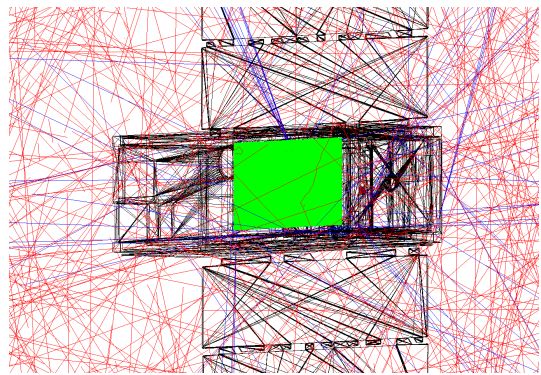}}
	\caption{Simulation of particles and photons interacting
with the mass model of the satellite in Geant4. The green rectangle marks the tracks of the optical photons produced in the scintillator of the detector.\label{fig:geant4}}
\end{figure}

\section{Short GRB spectrum}\label{sec:grb}
The main goal of the \textit{CAMELOT} mission is to detect and localize short GRBs, which are associated with mergers of neutron stars, and therefore strong sources of gravitational waves. Hence, the highest priority is to estimate the signal-to-noise ratio (SNR) expected from a typical short GRB.

The spectrum of a typical short GRB ($T_{90}<2$~s) was constructed by taking the typical values of the spectral pa\-ra\-me\-ters: peak energy $E_\textrm{peak}$, low energy spectral slope $\alpha$ and high energy spectral slope $\beta$ of the Band function \citep{Band1993} of short GRBs detected by \textit{Fermi}/GBM from the distributions published by \citet{Nava2011}. Then we tuned the normalization $A$ of the spectrum to obtain the values of the integral flux in the range 10-1000~keV equal to the median peak fluxes $F$ (ph\,cm$^{-2}$s$^{-1}$) of short GRBs observed by \textit{Fermi}/GBM in the same energy range and calculated from the Fermi GBM Burst Catalog (FERMIGBRST\footnote{\url{https://heasarc.gsfc.nasa.gov/W3Browse/fermi/fermigbrst.html}}), see \cite{Gruber2014, vonKienlin2014, NarayanaBhat2016ApJS}.

The spectral parameters of a typical short GRB used in the Geant4 simulations are $E_\textrm{peak}=500$~keV, $\alpha=-0.5$ and $\beta=-2.3$.
The median peak flux and the spectral normalization for 1024~ms time scale around a GRB peak are $F_{1024}=2.00$ ph\,cm$^{-2}$s$^{-1}$ and $A_{1024}=7.8\times10^{-3}$ ph\,cm$^{-2}$s$^{-1}$keV$^{-1}$, respectively.

\section{Monte Carlo simulations}\label{sec:simul}
Full Monte Carlo simulations including the satellite's mass model in Geant4 is performed (see Figure~\ref{fig:geant4}). Three different thicknesses of the Al casing, 1.0 mm, 1.5 mm and 2.0 mm are considered in the simulations to check the dependence of the count rates on the material thickness.

\begin{figure*}[t]
	\centerline{\includegraphics[width=0.5\linewidth]{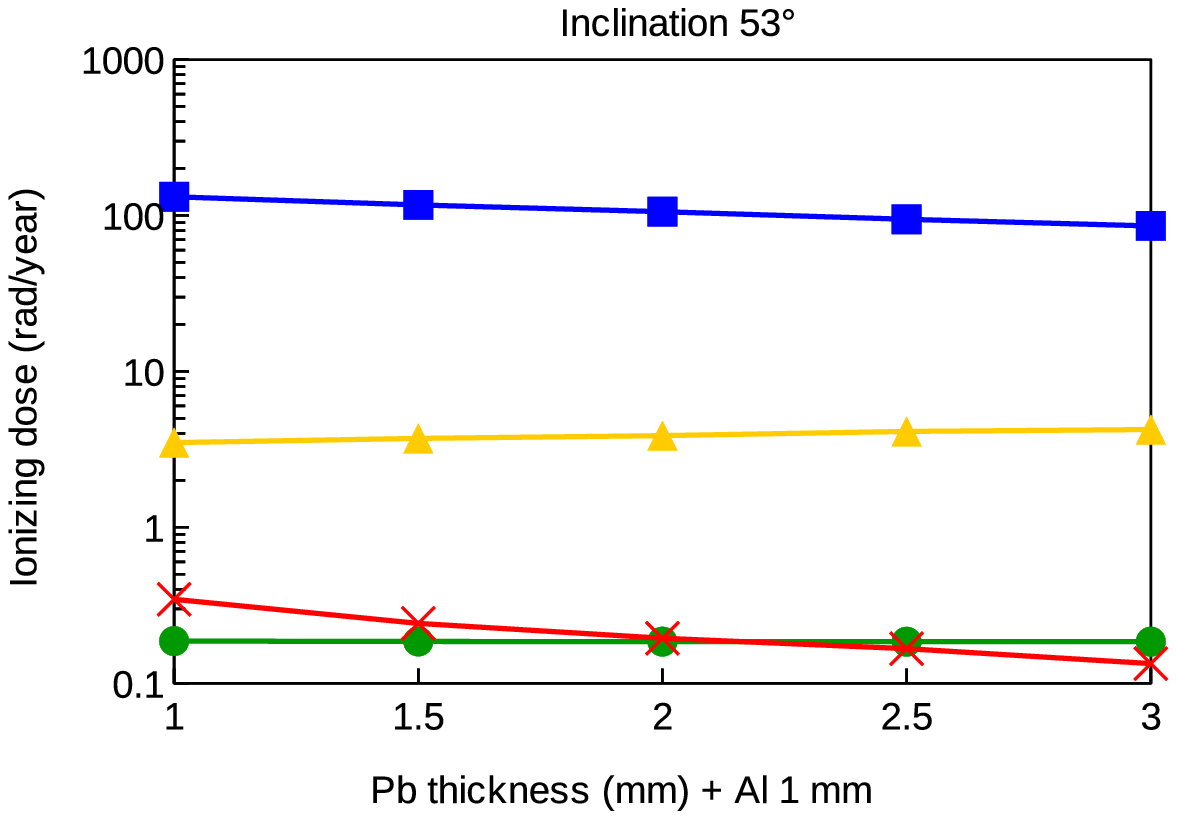}\includegraphics[width=0.5\linewidth]{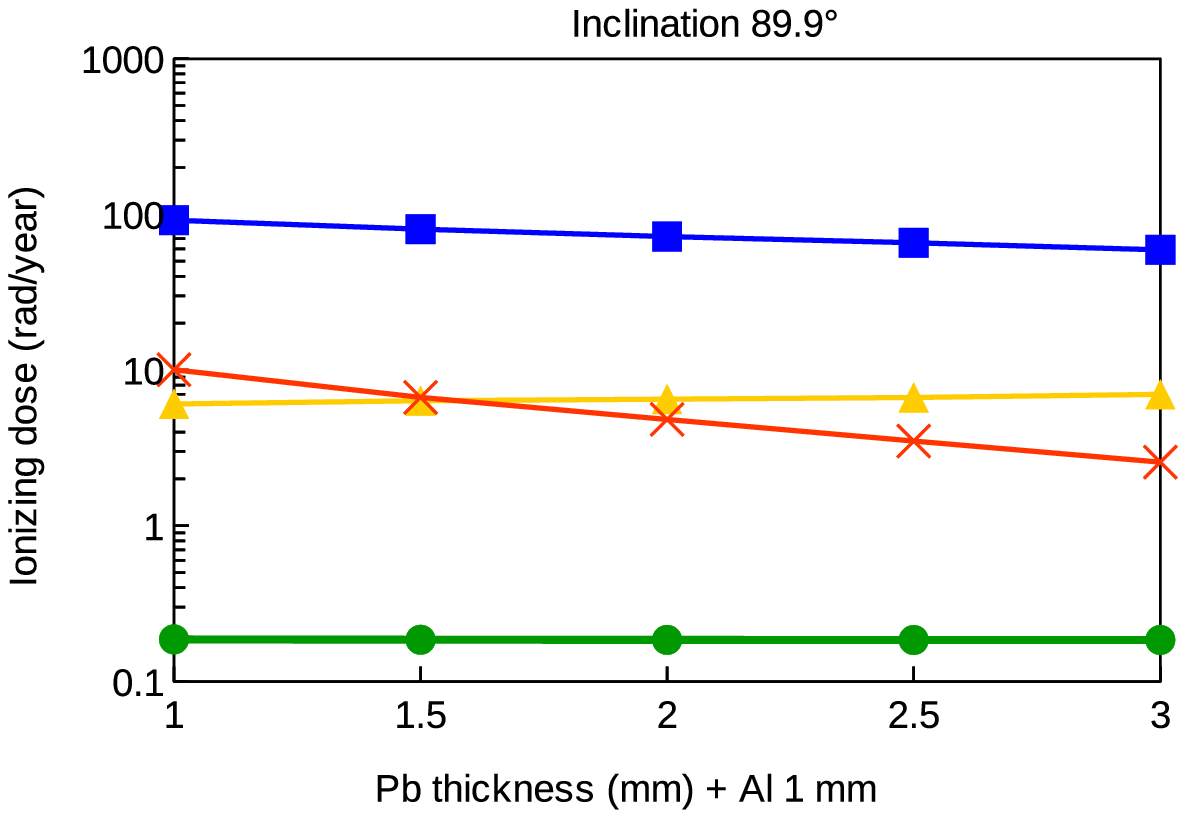}}
	\caption{Expected annual ionizing dose in MPPCs due to fluxes from various sources as a function of the thickness of the lead shielding. The simulated orbits have altitude 500~km, inclinations of $53^\circ$ (left) and $89.9^\circ$ (right). The sources are: trapped p$^+$ (blue squares), GCR H1+H2+He3+He4 (yellow triangles), solar p$^+$ (red crosses), secondary p$^+$ (green circles).\label{fig:dose}}
\end{figure*}

\subsection{Description of Geant4 simulations}
In order to simulate the interaction of particles including gamma-rays from GRBs and background particles, a dedicated particle physics simulation was developed using Geant4 10.0.p4. In order to be as realistic as possible -- to take into account the secondary particles created by the interaction of the primary particles and the material of the satellite too -- a complex CAD model of the satellite was imported to the Geant4 simulation with CADMESH \citep{Poole2012}. The ``Physics List'' used in the simulation was chosen to be a ModularPhysicsList that included elastic hadron scattering, all re\-le\-vant physics for neutrons, electromagnetic processes and optical photon physics.

Individual (scintillation) photons with an optical wavelength were created by particles depositing energy inside the scintillator. These were tracked until they were either absorbed or detected by the SiPMs implemented in the simulation. The optical parameters of the scintillators were determined by a dedicated set of measurements. The most relevant optical parameters were attenuation length, the reflectivity of the enhanced specular reflector (ESR) tape that is planned to cover the CsI(Tl) scintillators and the photon detection efficiency of the SiPM. The optical parameters were fine tuned by matching the measured and simulated spectra.

\subsection{Simulation results}
The detection count rates from the different components of the background obtained from the Geant4 simulations are presented in Table~\ref{tab:bkg_results}. The dominant component is CXB followed by the albedo X-rays and secondary positrons.

The Geant4 simulation for a typical short GRB with the incident flux perpendicular to the scintillator and not passing through the satellite's body gives detection count rates 135~cnt/s, 130~cnt/s and 125~cnt/s for Al scintillator casing thicknesses 1.0 mm, 1.5 mm and 2.0 mm, respectively. Assuming a trigger algorithm with a time window of 1~s then SNR = GRB counts / $\sqrt \textrm{background counts}$ would be 4.3, 4.8 and 5.1 for the three Al casing thicknesses, respectively.

The detection count rates were obtained for only one CsI scintillator of size $75\times150\times5$~mm$^3$. Up to four CsI scintillators (two scintillators on two perpendicular sides) are planned for each \textit{CAMELOT} 3U satellite.

\begin{center}
\begin{table*}[t]%
\caption{Detection background count rate (cnt/s) for one CsI(Tl) scintillator of size $75\times150\times5$~mm$^3$ as a function of the Al casing thickness obtained from the Geant4 simulations.
\label{tab:bkg_results}}
\centering
\begin{tabular*}{500pt}{@{\extracolsep\fill}lcccccccccccc@{\extracolsep\fill}}
\toprule
\textbf{Casing} & \multirow{2}{*}{\textbf{CXB}}  & \textbf{Albedo} & \multicolumn{4}{@{}c@{}}{\textbf{Secondary particles}} &  \multicolumn{4}{@{}c@{}}{\textbf{GCR}}  & \textbf{Trapped} & \multirow{2}{*}{\textbf{Total}} \\
\textbf{(mm)} & & \textbf{X-ray} & \textbf{p$^+$} & \textbf{n} & \textbf{e$^-$} & \textbf{e$^+$} & \textbf{H} & \textbf{He} & \textbf{e$^-$} & \textbf{e$^+$} & \textbf{e$^-$} & \\
\midrule
1.0 & 703 & 175 & 17 & 6.0 & 21 & 58 & 16 & 6.9 & 0.15 & 0.07 & 1.2 & 1007 \\
1.5 & 425 & 179 & 19 & 4.8 & 20 & 63 & 19 & 6.9 & 0.16 & 0.33 & 1.2 & 738 \\
2.0 & 324 & 164 & 18 & 5.1 & 21 & 63 & 19 & 7.0 & 0.15 & 0.33 & 1.2 & 622 \\
\bottomrule
\end{tabular*}
\end{table*}
\end{center}

\section{Duty cycle}\label{sec:duty}
As seen from Figure~\ref{fig:AE9} the flux of trapped e$^-$ can increase by several orders of magnitude in the polar regions or inside SAA compared to the flux near equator. The high increase of the background was also observed by the \textit{Lomonosov}/BDRG \citep{Svertilov2018} which is a gamma-ray transient detector sensitive in the energy range from 10~keV to 3~MeV. It consists of three modules of NaI(Tl) and CsI(Tl) scintillators covered by less than 1~mm thick Al window shield and read out by photomultiplier tubes. The instrument shows that background count rate can increase ~$50\times$ when passing SAA or polar regions.

Therefore we calculated the duty cycle for altitude 500~km and inclinations $53^\circ$ and $97.6^\circ$, which are the inclinations considered in the feasibility study of \textit{CAMELOT} \citep{Werner2018}. We simulated 1000 circular orbits at 500~km altitude and calculated the fraction of the time the satellite spend in the area with high background flux of trapped e$^-$ (> 1~cm$^{-2}$s$^{-1}$). For the map of the integral fluxes of trapped e$^-$ ($E>40$~keV) we used the AE-8 MIN Update ESA-SEE1 model \citep{Vampola1998}.

At 500~km altitude a satellite would spend 23\% and 32\% of the time in the regions of flux > 1~cm$^{-2}$s$^{-1}$ of trapped e$^-$ for inclinations $53^\circ$ and $97.6^\circ$, respectively.

\section{Expected ionizing dose in MPPC}\label{sec:dose}
It is considered to use Hamamatsu S14160-6050HS MPPCs for the detectors for \textit{CAMELOT}. The proton beam irradiation tests at W-MAST (Wakasa-wan, Japan) facility performed by the \textit{CAMELOT} team reveal that the MPPCs degrade when irradiated by 200~MeV p$^+$. Therefore it is necessary to estimate the expected total ionizing dose at (LEO) and design a protective cover shielding MPPCs to decrease the total dose due to p$^+$.

We used Multi-Layered Shielding Simulation (MULASSIS) \citep{Lei2002} implemented in SPENVIS to calculate ionization dose in a Si sphere shielded by Al and Pb. A simple spherical geometry with source flux isotropically incident over $4\pi$ srad was used for approximation.

The following proton sources are used in the simulation: trapped p$^+$ AP9 model (MC mode, 50\% CL, 100 runs, 30 days of orbit simulation with 60~s sampling); solar p$^+$ ESP-PSYCHIC \citep{Xapsos2007} model implemented in SPENVIS (50\% CL, magnetic shielding on, stormy magnetosphere, St\o{}rmer with eccentric dipole method, magnetic field moment unchanged); GCR p$^+$ ISO 15390 model (same parameter setting as described in Section~\ref{sec:GCR_H_He}); for secondary p$^+$ we use the same model as described in Section~\ref{sec:secondary_p}.

The expected annual doses for inclinations $53^\circ$ and $89.9^\circ$ are shown in Table~\ref{tab:dose} and in Figure~\ref{fig:dose}. The irradiation tests at W-MAST facility suggest that the considered MPPCs can be used for the \textit{CAMELOT} mission when a proper shielding is applied. For a 1U CubeSat demonstration mission GRBAlpha we consider to use 2.5~mm of Pb + 1.0~mm of Al.

\begin{center}
\begin{table*}[t]%
\caption{Expected ionizing dose (rad/year) in MPPCs for altitude 500~km as for different shielding.\label{tab:dose}}
\centering
\begin{tabular*}{500pt}{@{\extracolsep\fill}clcccccccc@{\extracolsep\fill}}
\toprule
 & \textbf{Shielding} & \multirow{2}{*}{\textbf{Trapped p$^+$}}  & \multirow{2}{*}{\textbf{Solar p$^+$}}  & \multicolumn{4}{@{}c@{}}{\textbf{GCR}}  & \multirow{2}{*}{\textbf{Secondary p$^+$}} & \multirow{2}{*}{\textbf{Total}} \\
 & \textbf{thickness (mm)} & & & \textbf{H1} & \textbf{H2} & \textbf{He3} & \textbf{He4} & &  \\
\midrule
\multirow{7}{*}{\rotatebox[origin=c]{90}{\textbf{Inclination 53$^\circ$}}} & no shielding & 1108  & 41  & 0.8 & 0.9  & 0.4  & 0.4 & 0.2 & 1152 \\
 & Al 1.0               & 223   & 1.2 & 0.9 & 1.1  & 0.5  & 0.5 & 0.2 & 227  \\
 & Pb 1.0    + Al 1.0   & 132   & 0.3 & 1.1 & 1.4  & 0.5  & 0.5 & 0.2 & 136  \\
 & Pb 1.5    + Al 1.0   & 117   & 0.2 & 1.2 & 1.5  & 0.5  & 0.5 & 0.2 & 121  \\
 & Pb 2.0    + Al 1.0   & 106   & 0.2 & 1.2 & 1.6  & 0.5  & 0.5 & 0.2 & 110  \\
 & Pb 2.5    + Al 1.0   & 95    & 0.2 & 1.3 & 1.7  & 0.5  & 0.6 & 0.2 & 99   \\
 & Pb 3.0    + Al 1.0   & 86    & 0.1 & 1.3 & 1.8  & 0.6  & 0.6 & 0.2 & 90   \\
\midrule
\multirow{7}{*}{\rotatebox[origin=c]{90}{\textbf{Inclination 89.9$^\circ$}}} & no shielding & 773  & 1262  & 1.6 & 1.8  & 0.8  & 0.8 & 0.2  & 2040 \\
 & Al 1.0               & 153  & 36    & 1.7 & 2.0  & 0.8  & 0.8 & 0.2  & 194  \\
 & Pb 1.0    + Al 1.0   & 92   & 10    & 2.0 & 2.4  & 0.8  & 0.9 & 0.2  & 108  \\
 & Pb 1.5    + Al 1.0   & 80   & 6.7   & 2.1 & 2.6  & 0.9  & 0.9 & 0.2  & 94   \\
 & Pb 2.0    + Al 1.0   & 72   & 4.8   & 2.1 & 2.6  & 0.9  & 0.9 & 0.2  & 83   \\
 & Pb 2.5    + Al 1.0   & 66   & 3.5   & 2.2 & 2.7  & 0.9  & 0.9 & 0.2  & 76   \\
 & Pb 3.0    + Al 1.0   & 59   & 2.6   & 2.3 & 2.9  & 0.9  & 0.9 & 0.2  & 69   \\
\bottomrule
\end{tabular*}
\end{table*}
\end{center}

\section{Conclusions}\label{sec:conclude}
We studied the expected background for the \textit{CAMELOT} mission and detectability of short GRBs. It was shown that with at least one large scintillator short GRBs with a typical peak flux can be detected. Future work will include, for example, a study of the background activation due to the radioactive decay of elements of the detector and the satellite's body ge\-ne\-ra\-ted when passing SAA. Also other transients, beside the short GRBs, such as long GRBs, SGRs, TGFs should be considered. The expected annual doses allow to use the planned MPPCs for the \textit{CAMELOT} mission when a proper shielding against protons is applied.

\section*{Acknowledgments}

This research has been supported by the grants awarded by the \fundingAgency{Hun\-ga\-rian Academy of Sciences}, Lend\"ulet \fundingNumber{LP2016-11, KEP-7/2018}, and by the European Union, co-financed by the \fundingAgency{European Social Fund} (Research and development activities at the E\"{o}tv\"{o}s Lor\'{a}nd University's Campus in Szombathely), \fundingNumber{EFOP-3.6.1-16-2016-00023}.


\subsection*{Financial disclosure}
None reported.

\subsection*{Conflict of interest}
The authors declare no potential conflict of interests.

\bibliography{ripa_ibws2019}%

\section*{Author Biography}

\begin{biography}{\includegraphics[width=60pt,height=70pt]{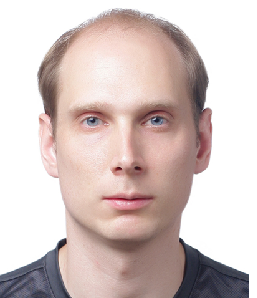}}{\textbf{Jakub {\v R}{\'i}pa} received the Master's degree in physics in 2006 and the Ph.D. degree in astronomy and astrophysics in 2011, both at the Charles University (MFF UK), Prague, Czech Republic under the supervision of doc. RNDr. Attila M\'{e}sz\'{a}ros, DrSc. He was a postdoctoral researcher at: Institute for the Early Universe, Ewha Womans University, Seoul, Korea (2011-2012); Institute of Basic Science, Sungkyunkwan University, Suwon, Korea (2012-2015); Leung Center for Cosmology and Particle Astrophysics, National Taiwan University, Taipei, Taiwan (2015-2017). Currently, he is a postdoctoral researcher at the Astronomical Institute, Charles University and at the Lend\"ulet Hot Universe Research Group, MTA-E\"ot\"vos University, Budapest, Hungary. His research interests include gamma-ray bursts and instrumentation for high-energy astrophysics in hard X-rays and gamma-rays.}
\end{biography}

\end{document}